\documentclass[preprint,12pt,3p]{elsarticle}

\usepackage{tabularx, booktabs}
\newcolumntype{Y}{>{\centering\arraybackslash}X}

\usepackage{lineno}
\usepackage{times}
\usepackage{graphicx}
\usepackage{gensymb}
\usepackage{array}
\usepackage{amssymb}
\def\parencite{\citep}
\def\textcite{\citet}
\usepackage[fleqn]{amsmath}
\usepackage{amsthm}
\usepackage{rotating}
\usepackage{caption}
\newtheorem*{and et al.*}{Definitia}

\biboptions{round, authoryear}

\journal{Icarus}

\begin{document}

\begin{frontmatter}

\title{Relevance of Tidal Heating on Large TNOs}
%\tnotetext[label0]{This is only an example}

\author[label1]{Prabal Saxena\corref{cor1}}
\address[label1]{NASA/Goddard Space Flight Center,\\8800 Greenbelt Rd, Greenbelt, MD 20771, USA}

\cortext[cor1]{Corresponding author}

\ead{prabal.saxena@nasa.gov}

\author[label2]{Joe Renaud}
\address[label2]{Department of Physics \& Astronomy, George Mason University\\4400 University Drive, Fairfax, VA 22030, USA}

\author[label1,label3]{Wade G. Henning}
\address[label3]{Department of Astronomy, University of Maryland\\Physical Sciences Complex, College Park, MD 20742}

\author[label4]{Martin Jutzi}
\address[label4]{Physics Institute: Space Research \& Planetary Sciences, University of Bern\\ Sidlerstrasse 5, 3012 Bern, Switzerland}

\author[label1]{Terry Hurford}

\begin{abstract}
We examine the relevance of tidal heating for large Trans-Neptunian Objects, with a focus on its potential to melt and maintain layers of subsurface liquid water.  Depending on their past orbital evolution, tidal heating may be an important part of the heat budget for a number of discovered and hypothetical TNO systems and may enable formation of, and increased access to, subsurface liquid water. Tidal heating induced by the process of despinning is found to be particularly able to compete with heating due to radionuclide decay in a number of different scenarios. In cases where radiogenic heating alone may establish subsurface conditions for liquid water, we focus on the extent by which tidal activity lifts the depth of such conditions closer to the surface. While it is common for strong tidal heating and long lived tides to be mutually exclusive, we find this is not always the case, and highlight when these two traits occur together.      
\end{abstract}


\end{frontmatter}

% \linenumbers
\normalsize
\section{Introduction}
\label{sec1}

It appears highly likely that numerous bodies outside the Earth possess global subsurface oceans - these include Europa, Ganymede, Callisto, Enceladus \citep{1999JGR...10424015P, 1998Natur.395..777K, Thomas201637, JGRA:JGRA51618}, and now Pluto \parencite{Nimmo2016}. Other bodies may also have subsurface water such as Triton \parencite{2007ess..book..483M} and Dione \parencite{GRL:GRL55051}. Numerous other moons may have possessed significant subsurface liquid water in the past. Perhaps most surprisingly, even bodies far beyond the snow line that do not orbit a major planet have hinted at surface and subsurface liquid water. Besides Pluto, Quaoar, Haumea and (225088) 2007 OR10 all have exhibited either bulk density, surface color, and$/$or spectroscopic evidence that may be consistent with surface liquid water activity \parencite{2004Natur.432..731J, 2007ApJ...655.1172T, 2011ApJ...738L..26B}. 

Other Trans-Neptunian Objects (TNOs) have also exhibited remarkable diversity with respect to evidence of resurfacing and water absorption features. Given estimates of over 100,000 TNOs with radii greater than 100 km in the main classical Kuiper belt alone \parencite{2011AJ....142..131P}, such evidence of activity on already detected bodies portends a potentially vast reservoir of liquid water. The observational evidence on some of these TNOs that may suggest the presence of subsurface water is in some cases supported by theoretical modeling (as an example, see \textcite{HUSSMANN2006}). However, the observational evidence considered a potential proxy for interior water is not entirely conclusive. Spectroscopic evidence of crystalline water ice or ammonia hydrates may be due to cryovolcanism. However, micrometeorite impacts may anneal surface layers and expose crystalline water ice \parencite{Porter2010492}. Ammonia hydrates may be a result of diffusion of ammonia from subsurface layers that are shielded from irradiation \parencite{Cruikshank201582}. Additionally, interior modeling of TNOs is sensitive to the inclusion of different heat sources as well as assumptions regarding the interior structure and material properties. In previous models including \textcite{HUSSMANN2006}, a number of larger TNOs are right on the edge of being able to host subsurface liquid layers. These models, for the most part, only treat radiogenic heating fully, as it is assumed to dominate. However, additional sources of heat may provide important modifications to their thermal history even if they are lesser in magnitude or more short-lived.

The dominant source of heat for TNOs is radiogenic heating. Early thermal evolution of some outer solar system bodies may have been strongly influenced by short-lived isotopes such as $^{26}$Al and $^{60}$Fe, with some potential to initiate differentiation \parencite{2008ssbn.book..213M,Bizzarro2005}. Long-lived isotopes $^{238}$U, $^{235}$U, $^{232}$Th and $^{40}$K, each contribute to interior and surface evolution over solar system timescales. Antifreeze solutions may reduce the melting temperature of a subsurface water brine. Ammonia (eutectic point $\sim$176 K for a water-ammonia
mixture at 1 bar \parencite{CROFT1988279}) is considered to be the most likely antifreeze in potential TNO subsurface oceans \parencite{2002Icar..159..518L, 2005P&SS...53..371G}, with methanol (eutectic point $\sim$157 K for a water-methanol
mixture \parencite{doi10.1021/je60022a017}) and a variety of salts also possible candidates. An ammonia mixture with water ice may also make an ice shell more conductive and less viscous \parencite{Robuchon2011426}. 

Beyond radiogenic heating, other heat sources include leftover accretion heat, latent heat release, tidal heat, and ongoing differentiation heat. While radiogenic heating is the largest reservoir of heat available to Pluto (distributed over the longest period of time), after accretional heat, tidal heating due to despinning may be the next largest reservoir (See e.g., \textcite{Robuchon2011426} table 2). Total ongoing accretional heat escape is often comparable to radiogenic heat on Earth-sized bodies but rapidly lost in small bodies due to shallow burial during assembly \parencite{Robuchon2011426}. Further, Pluto studies find tidal despin heating $<$2$\times$ smaller than accretion heating and greater than both differentiation and latent heat \parencite{HUSSMANN2006, 2016GeoRL..43.6775H}. Using initial Charon eccentricity values from formation cases in \textcite{Canup2005} can result in eccentricity-tide heating far larger ($>$10$\times$) than the spin-tide heating cases above. \textcite{Cheng2014242} also show Pluto's tidal evolution may take $\sim$10$^{7}$--10$^{8}$ years and may include temporary capture into a range of resonances, implying much work remains for this system. 

An accurate picture of potential tidal heating requires modeling of initial orbital and bulk properties after a system-forming collision. The importance of this collision modeling is underpinned by numerous observations which suggest many large TNOs have undergone a major collision in their history. Some of the largest objects that bear the most distinct signs of geologic activity, or water, are also bodies which appear modified in the past by a significant collision (Pluto, Haumea and Varuna \parencite{2015Sci...350.1815S, 2009A&A...496..547P, 2002AJ....123.2110J}). A collisional history is not necessarily restricted to just these bodies - collisionally altered worlds may have had their surfaces converted back to cold, primordial surfaces over time. Additionally, many of the largest TNOs have only been discovered in the last 20 years, and it is likely that other similar bodies will be discovered \parencite{1538-3881-151-2-22}. Given the importance of collisions in shaping the outer solar system \parencite{2008AJ....136.1079L}, their role in relation to the initial orbital parameters and thermal state of TNO systems needs to be included. 

This study examines the importance of tidal heating for large TNOs in order to determine their potential influence on the creation and persistence of subsurface water on these bodies.  Tidal heating may occur due to 1.) non-zero eccentricity, 2.) non-synchronous spin, and 3.) non-zero obliquity. Tidal dissipation drives systems towards equilibrium states, in part by evolving eccentricity and obliquity to zero, and spin rates toward synchronization, all at rates inversely proportional to the intensity of dissipation. There is a perception that this means spin and obliquity tides are intense but brief, and eccentricity tides, if large enough to be meaningful, are also short-lived (in the absence of a locked resonance such as for Europa). For these reasons many expect the TNO region tidally quiescent. However, there is a paucity of tidal modeling of TNO systems from formation through orbital evolution, particularly given newer more accurate rheological models. Given the thermal state of TNOs that are predicted to exist based upon radiogenic modeling alone, we examine the potential for tidal heating to act as a tipping point for some of the largest TNOs. These TNOs are plausible targets for future observation, and understanding their thermal evolution more fully could enable better interpretation of surface features - particularly any involving past or extant liquid water.

\section{Classical Tidal Outcomes}
\label{sec2}

To develop a portrait of tidal relevance for the TNO population, in Table 1 we use the classical tidal heating equation for arbitrary eccentricity and obliquity \parencite{Wisdom2008} to estimate possible post-collisional conditions for known TNO binaries. Spin tides are calculated using the classic tidal spin acceleration formula found in e.g., \textcite{Goldreich1966}, which may be converted into a formula analogous to the standard tidal dissipation formula for eccentricity tides,

\begin{equation}
    \dot{E}_{spin} = \Omega \frac{3}{2} \frac{k_2}{Q} \frac{G R_{pri}^5 M_{sec}^2 }{a^6}
\end{equation}

where $\dot{E}_{spin}$ is the dissipation rate, $\Omega$ the spin rate, $k_2$ the second order Love number, $Q$ the tidal quality factor, $G$ the gravitational constant, $R_{pri}$ the radius of the object being despun, $M_{sec}$ the mass of the perturber, and $a$ the semi-major axis. The same $\dot{E}_{spin}$ formula may also be derived from formulations such as \textcite{CastilloEfroimskyLainey2011} eqn 44a. We use the obliquity damping timescale from \textcite{BillsNimmo2011} eqn 65, and despinning timescale $\tau_s$ from \textcite{MurrayDermott2005} exercise 5.2, which is adapted to handle cases where the secondary mass is substantial in comparison to the primary mass. Note that the $\tau_{obl}$ formula used is actually adapted from a timescale for free axis wobble damping \parencite{BillsNimmo2011}. For simplicity, and to make cross-object comparisons mainly based on mass and separation, results in Table 1 use a fixed frequency-insensitive value of $Q$=100. This is revisited in section \ref{sec3}. Love numbers $k_2$ are computed using a rigidity 4$\times$10$^9$ Pa. Unlike for larger gravity dominated planets $\geq$ $\sim$ M$_{Mars}$, $k_2$ values for TNOs vary strongly by mass and compete with $Q$ for principal control of tidal interior outcomes. We also find by the methods of \textcite{HenningHurford2014} that inhomogeneous viscoelastic $k_2$ values can be up to 10--30$\times$ higher, especially in the presence of an ocean, with corresponding (linear) elevations in $\dot{E}$ and decreases in $\tau$.  

Upper limits on post-collisional spin are set by the pull-apart strength limit, which is a function only of density \parencite{CanupRighter2000}.  We test values up to 90\% this rate for each object. When unknown, densities of 1.5 g cm$^{-3}$ are applied, leading to a 90\% spin limit of 3 hours.  

We assume the silicate portion of TNOs have radionuclide heat production rates equal to the estimated Earth lower mantle rate $\sim$1.5$\times$10$^{-8}$ W/m$^3$ \parencite{Jaupart2007}. Some hypothetical TNOs may have far lower silicate fractions (such as for comets), which would mean tides dominating radionuclides with greater ease. Layer structures are modeled assuming simple densities of 1000, 3000, and 7000 kg/m$^3$ for water/ice, silicate, and Fe/Ni respectively. Results are not sensitive to whether or not Fe/Ni has fully separated from the silicate, and we assume a 30\% (roughly chondritic) silicate to iron mass ratio. For this analysis it is sufficient to leave the density of liquid and solid water equal.   

We calculate the term $\Delta{}D$, or the change in the depth of habitable temperatures, relative to a nominal habitable depth $D$ with radionuclides alone, as determined for a conductive heat solution (using a thermal conductivity of 2.1 W K$^{-1}$ m$^{-2}$). Values of $D \geq R$ (marked with $^\dagger$) mean radionuclides alone fail to generate habitable conditions anywhere in the object. Large values of $\Delta{}D$ mean tides, at least initially, lead to an important upward shift in such a habitable horizon.   

Table 1 shows numerous cases (in bold) exist with long lived tides ($\geq$ 1 Gyr), high initial magnitude tides ($\geq$ 5\% of $\dot{E}_{Rad}$), and a few cases in between that exhibit both. Not surprisingly, spin tides generate the highest intensity heating, with eccentricity tides second, and obliquity tides least. Spin tides damp much slower than obliquity tides, allowing for marginal overturn of the conventional notion that both spin and obliquity tides both damp rapidly. This can be explained by the fact that the $\tau_{spin}$ equation actually contains the spin rate $\Omega_{initial}$, whereas $\tau_{circ}$ does not contain $e_{initial}$, and $\tau_{obl}$ does not contain the initial obliquity angle. This difference is critical, and allows spin tide damping times to scale in a rate-dependent manner unlike other tides. For initial spin rates in the 10 h range this would hardly matter, but given post-collision spin rates up to the pull-apart limit of $\sim$2--3 h are allowed, this creates a category of tidal outcomes that are both meaningful in intensity as well as durable. 

Three hypothetical system cases are included in Table 1, with dramatic pairings of tidal intensity and duration co-occurring for modestly greater masses or semi-major axis adjustments. Ongoing discoveries suggest such perturbations on Eris-Dysnomia or Pluto-Charon system parameters likely reside among the undiscovered TNO population.  

Ratios of tidal heating to radiogenic heating in Table 1 (referred to as the Heat Rate Ratio, or HRR) often exceed 1.0, indicating that each of the tidal mechanisms, as well as their summed effect, may easily dominate interior evolution for post-collisional states. It remains only a question of what fraction of collisions end with orbital/spin parameters at these levels (a question hydrodynamic modeling \parencite{2008Icar..198..242J, 2015Sci...348.1355J} will eventually answer). Note we are not actually testing the maximum plausible values for $e_o$ and $\Omega_o$, but this is still sufficient to demonstrate the point that there is a powerful role for tides in altering the subsurface water state of TNOs over time.   

Shifts in the `habitable horizon' depth $\Delta{}D$ due to tides are shown in the next to last column of Table 1, and demonstrate that tidal activity for such original states often causes important upwards shifts in the melting horizon for even a simple non-eutectic model of melting. Unlike $D$ itself, a useful feature of using $\Delta{}D$ as a metric is that it will be the same regardless of the melting temperature (brine composition) considered. While some significant $\Delta{}D$ values occur in the Orcus-Vanth system, they remain insufficient to achieve melting inside the objects for a 273 K threshold.

No compelling case for a tidal ocean can be made for 2007-OR10 or its satellite. The masses and periods of this system are very poorly constrained \parencite{Schwamb2009}, yet we fail to generate notable tides even with best-case parameters. Radionuclides however are able to allow liquid water in an undifferentiated 1.5 g cm$^{-3}$ model prior to the object center. With the possible exception of earlier tidal stresses that may have aided fracturing, observations of methane and water signatures are likely due to the previously noted hypotheses of a radionuclide driven subsurface ocean and the capability of the planet to retain surface methane \citep{2011ApJ...738L..26B}.

The most notable result in Table 1 is for Eris perturbed by Dysnomia starting at $\Omega$ = 2.3 hours. In this case a radionuclide-only ice shell of $D$ = 154 km may be reduced by 24 km or $\sim$15\%, and does so for longer than the solar system age. Adjusting $k_2$ to 0.17, appropriate for an ice fraction of 0.5, thins the shell by 40 km, or a 26\% reduction, enduring for 8 Gyrs. Lastly, testing $Q$ = 20 (plausible given Europa's similar response) in this case leads to $\Delta{}D$ = 98 km, or a 64\% reduction to 56 km total, lasting 1.7 Gyrs. These values assume a 273 K melting point.   

Key lessons from these classical tests are as follows: Obliquity tides are almost never high enough in $\dot{E}$ or $\tau$ to be significant. This further implies ongoing Cassini-state nonzero obliquity dissipation, while perhaps durable, would rarely compete with radionuclide heat. Eccentricity tides can be intense, but are generally short lived except in the Eris-Dysnomia system. Blend cases are mainly dominated by spin-tide solutions. Spin tide solutions are a key area for both high $\dot{E}$ \& $\tau$, due to the scaling of $\tau_s$ linearly with initial $\Omega$, and the high initial rates allowed from collision modeling. Objects with high mass in the object being despun have optimal $\dot{E}$ \& $\tau$ properties, due to simple inertia.

\section{Viscoelastic Results}
\label{sec3}

Viscoelastic results as functions of eccentricity and spin rate are shown in figure \ref{fig:spin_eccen}. These results are computed assuming thermal equilibrium between tidal heat production and escape. Parameterized convection is assumed to transport heat out of the silicate core. The convective vigor is strongly sensitive to viscosity \parencite{Henning2009} and becomes very ineffective below $\approx 1000$K. 

For simplicity, we only consider a conductive ice shell wherein we assume a bottom percentage to be in a viscoelastic regime that responds strongly to the tidal forcing frequency. The viscoelastic ice represents the locale of maximum tidal dissipation within the ice shell. \textit{Maximum liquid} percentage is defined using the thickness between the base of the thermally conducting ice and the top of the silicate core. A convecting ice layer may or may not be present, depending on the ice shell's history. The additional complexities required to compute the minimum liquid water layer thickness in a model with both conductive and convective ice is reserved for future work. Results using both the Maxwell \parencite[e.g.,][] {Segatz1988, FischerSpohn1990, HUSSMANN2006} and Andrade \parencite{CastilloEfroimskyLainey2011, Efroimsky2012, Shoji2013, EfroimskyMakarov2014, Kuchta2015} rheologies are shown, where Andrade results represent improved frequency domain sensitivity that better corresponds to laboratory experiments \parencite[e.g.,][]{Raj1971, Gribb1998, Faul2015}. To our knowledge this is the first Andrade application for TNOs. The Andrade rheology notably increases tidal dissipation at low temperatures (relative to the Maxwell response), and this can either extend or shorten tidal durations depending on intensity. In all results we assume the same rheology type for both the viscoelastic ice shell and the silicate core. The core is subjected to additional partial melting tracking \parencite{Moore2003b}. This, however, weakly affects TNO's whose core temperatures typically stay well below the silicate solidus.

\subsection{Time-Domain Results}
\label{sec3.1}

Both sections \ref{sec3} and \ref{sec2} look at static pictures of TNO systems. In reality any changes to a primary or satellite's spin or eccentricity will affect the system's semi-major axis. Since tidal dissipation (for all types of tides) is a strong function of semi-major axis, we find it is critical to take the role of a variable semi-major axis into account \parencite[e.g.,][]{Darwin1880, Ferraz-Mello2008}. Figure \ref{fig:tno_time_domain} shows full time histories for systems with a Pluto \& Charon analog pair, a Pluto-analog with reduced melting temperatures due to the presence of NH$_{3}$, and a system wherein the masses of both Pluto and Charon are reduced by half. The time domain runs from a post-collisional state with both a high spin rate and high eccentricity at $t=0$. This $t=0$ could be considered close to solar system formation for a conservative case, but may be much later. The initial silicate core temperatures are found by radiogenics alone. Orbits are evolved in time using classical equations for $de/dt$, $da/dt$, and $d\Omega/dt$ \parencite{Barnes2008, HUSSMANN2006} without resonances. We use a simplified switching mechanism that modifies these formulae based on the relative values of Pluto's spin ($\Omega_{pluto}$) and orbital motion ($n$). When $\Omega_{pluto} >> n$, its spin increases the semi-major axis ($a$). Once $\Omega_{pluto}$ has dampened ($\Omega_{pluto} \approx n$), this effect is overcome by any non-zero eccentricity dampening which will tend to pull back in $a$ \parencite{Ferraz-Mello2008}. Eccentricity tide viscoelastic responses (through $\text{Im}(k_{2})$) are modeled in both the Pluto and Charon analogs. Spin tides are only tracked in the Pluto analog. The Charon analog is assumed to have fallen into a 1:1 resonance with $n$ very quickly after formation. Initial radionuclide rates in the silicate core of each body are assumed that evolve in time towards modern chondritic rates.

Tidal heat is computed following \textcite{Henning2009} using the homogeneous tidal equation with a term to adjust for the volume of material actually contributing to tidal action. This term is constant for dissipation within the silicate core, but varies for the viscoelastic ice as its thickness changes. We assume no dissipation in the liquid H$_{2}$O layer, an assumption that may be revisited in the future based on suggestions that fluid heating, especially for thin liquid oceans, may have an important role \parencite{TylerHenningHamilton2015}. Multilayer results \parencite{HenningHurford2014} demonstrate that the sub-layer of any asthenosphere or ice shell that is viscoelastically best tuned to a given forcing frequency will overwhelmingly dominate the tidal response. For these TNO models, that viscoelastic layer is the lowest $\sim$10\% of the conducting ice shell, where ice is warmest. We validate this assumption with a small set of full multi-layer solutions for static cases, prior to using the classical method with a volume adjustment for these simulations. This approximation is justified for the purpose of assessing an order of magnitude sense of plausible tidal outcomes and duration. Solutions are far more sensitive to the selection of ice viscosity, or by proxy grain size. We use a grain size of 0.001 mm, and the ice rheology law of \textcite{GoldsbyKohlstedt2001} with parameters from \textcite{Moore2006}. 

The range of final $a$ in figure \ref{fig:tno_time_domain} does not reproduce present day measurements. This is due to the spin quickly pushing out $a$ while it is synchronizing with $n$. Eccentricity is able to pull this $a$ back in, however, at the relevant points in time, it too has been nearly lost. For simplicity of interpretation, we choose to display fixed start cases instead of fixed end states, but the spread in outcomes is what matters (Thus these represent 1M$_{Pluto}$-analogs, not Pluto itself). Spin tides due to a fast spinning primary typically cause $a$ to expand at first, but then later contract in phases after primary spin synchronization is expected. We find, using $\Delta{}D$ as our metric, that tides can increase liquid water volumes over radiogenics alone for nearly the age of the solar system in these limited test cases. Our conclusion from these experiments is that TNO binary tidal evolution is well within the realm necessary to have near-past or present day ocean worlds within the TNO belt assisted by tides, but that considerable future analysis is warranted. 

% Joe, this is meant just to be an initial template paragraph based on your data so far, to help jump start your description of findings. Alter at will! - WH June 3rd
In developing figure \ref{fig:tno_time_domain}, we find severe sensitivity of system outcomes to the starting semi-major axis $a_o$. Small $a_o$ leads to rapid tidal evolution, rapid outward movement to larger $a$ (due to high primary spin rates), and only a short ($\leq$50-100 Myr) episode where tides approach the order of magnitude of radiogenics. Large $a_o$ leads to long tidal durations, but negligible magnitudes. The Goldlocks region in $a_o$ to achieve meaningful tidal magnitudes and duration's appears to lie between 1.5 and 2.5 $a_{present}$for a Pluto-Charon analog system. A more thorough investigation of $a_{o}$, $\Omega_{o}$, and $e_{o}$ values, preferably including the role of resonances, will be required to find the circumstances for specific real systems that can lead to large, long-lasting tides that also produce realistic end conditions that we find today.

\begin{figure}[t]
\centering
\includegraphics[width=0.8\textwidth]{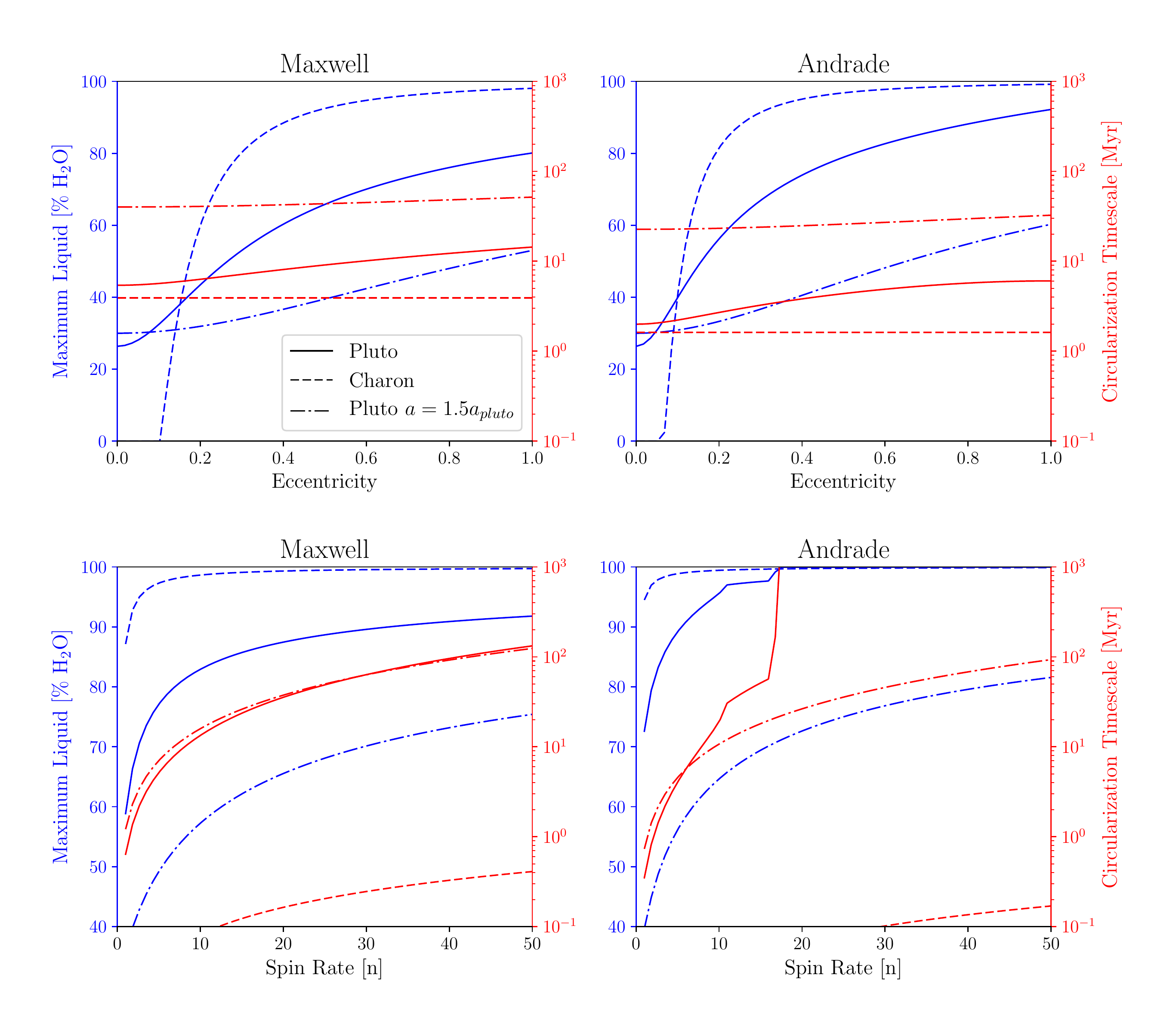}
\caption{Pluto and Charon analogues are subjected to forced eccentricities (top row) and spin rates (bottom row). Present day Pluto and Charon are shown (solid and dashed respectively), along with Pluto starting at $1.5\times$ its current semi-major axis (dashed-dot). Equilibrium liquid H$_{2}$O depth percentages are found at each eccentricity or spin. Heating is calculated as a summation of fixed radiogenic rate comparable to present-day chondritic isotope concentrations and the respective eccentricity/spin tidal dissipation. The maximum liquid water percentage (blue left y-axis) is calculated by finding the equilibrium thickness of the elastic+viscoelastic ice shell thickness. Radiogenics alone produce non-trivial amounts of liquid water on Pluto no matter the tidal interactions. However, tides (especially those produced by spins) can greatly enhance liquid water growth. Damping timescales (defined as $\dot{x}/x$ where $x$ is either eccentricity or spin) is shown in red on the right y-axis. Note that both the liquid H$_{2}$O percentages and timescales are strongly coupled to orbital motion which remains fixed in these results. The effects of a varying orbital motion are explored in section \ref{sec3.1}. The rapid jumps seen in some Andrade Rheology results are due to silicate core convection balancing heat production in such a way that equilibria can jump with very small changes. This interplay will be explored in a future study.}
\label{fig:spin_eccen}

\end{figure}

\begin{figure}
\centering
\includegraphics[width=0.8\textwidth]{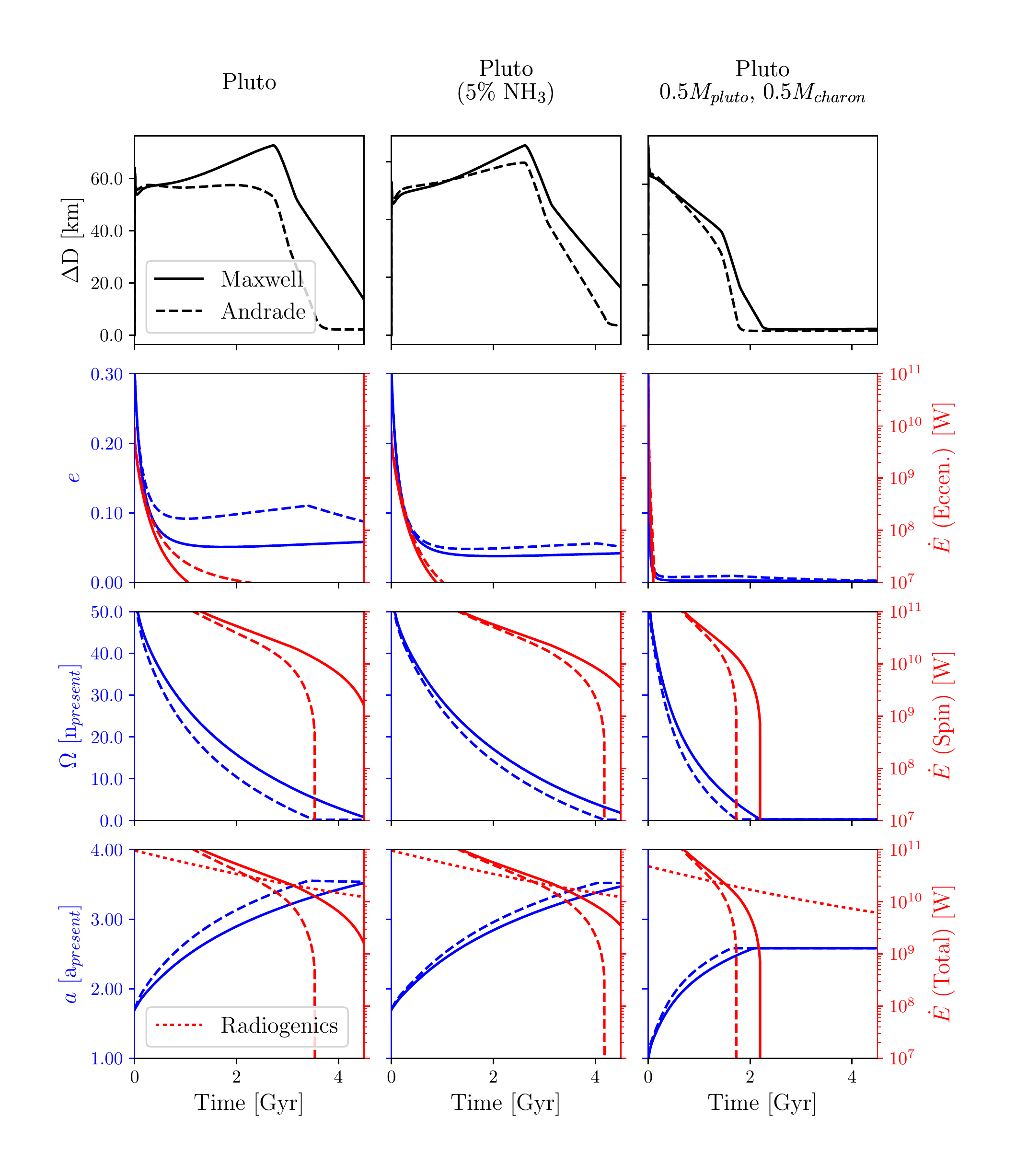}
\caption{Time histories for a simplified thermal-orbital-spin evolution on a Pluto-analog (col 1), a Pluto-analog with $\approx5\%$ NH$_{3}$ mix of H$_{2}$O (col 2), and a reduced mass Pluto-Charon system (both with half their present mass) (col 3). Initial conditions mimic post-collision orbital/spin parameters discussed in section \ref{sec1}. Initial silicate core temperatures are calculated to equilibrate with radiogenics alone at $t$=0. Eccentricity and non-synchronous spin are lost due to tidal heating (rows 2 \& 3). Likewise, distance between primary and secondary is also modified by the systems' angular configurations, with $a$ tending to increase when $\Omega > n$ and decrease when $\Omega \approx n$ and $e \gg 0$ (row 4), discussed further in section \ref{sec3.1}. The habitable horizon ($\Delta{}D$, row 1) measures the increase in ocean thickness that both spin \& eccentricity tides cause over radiogenics alone. This thickness is found by monitoring the growth/loss of the purely conductive ice lid \parencite{Hussmann2004}.}
\label{fig:tno_time_domain}
\end{figure}

To explore the role of slight variations of planetary parameters, we also examine a reduced semi-major axis in figure \ref{fig:spin_eccen}, and a reduced mass system in the last column of figure \ref{fig:tno_time_domain} -- all keeping other parameters consistent with Pluto/Charon for context. Increasing the semi-major axis, unsurprisingly given the time-domain results presented here, decreases tidal importance while increasing longevity. Reduced mass results led to tides dominating radiogenics, but for shorter time periods (see figure \ref{fig:spin_eccen}). 

Lastly, let us briefly note a complex interplay can occur for cold-start silicate mantle conditions. If we step away from an assumption that the mantle is in full thermal equilibrium, with convection already occurring at $t$=0, then a new pattern in time occurs. TNO total radionuclide heat rates are sufficiently weak that the temperatures required for silicate convection to initiate do not occur for 1-2 Gyrs. The exact delay depends on the cold start temperature assumption, which is akin to estimating the thermal efficiency during accretion of trapping/burying gravitational potential heat. Once silicate convection does begin, radiogenic heat switches roles from at first merely warming the mantle (with minimal expression upon the ice shell), to escaping into the ice layer and then deep space in full equilibrium thereafter. This phenomenon can open a unique niche for tidal heating in the intervening 1-2 Gyrs, as tides can be the only significant heat source determining ice shell structure in this time period. Whether or not a given TNO has a hot or cold start depends mainly on when an assumed collisional start for the tidal dynamics occurs relative to gross accretional assembly of the TNOs involved. Collisions/dynamically start-points 4.5 Gyrs ago should be predominantly cold-start mantles, while events 1-2 Gyrs into solar system history will be mainly warm-start. 

\noindent\makebox[\textwidth]{
\small
\begin{tabularx}{\textwidth}{|c|c|c|c|c|c|c!{\vrule width 2pt}c!{\vrule width 2pt}c|c|c|c|}
\hline
\cline{1-12}
TNO Pair
& $R$
& Period
& $\rho$
& $k_2$
& $\dot{E}_{Rad}$
& $D$
& Case
& $\dot{E}_{Tidal}$
& HRR
& $\Delta$D
& $\tau$
\\
primary$/$perturber
& (km) 
& (days) 
& (g$/$cc) 
&
& (W) 
& (km) 
& $ e/ \Omega/ \phi $ 
& (W) 
&  
& (km) 
& (Gyr)
\\
\hline
Eris/Dysnomia & 1163 & 15.77 & 2.5 & 0.089 & 5.4e10 & 154
& $e$=0.2 & 1.7e7 & 0.0003 & 0.05 & 0.96\\
\hline
&&&&&&
& $e$=0.8 & 2.7e8 & 0.005 & 0.77 & 0.96\\
\hline
&&&&&&
& $\Omega$=4.6h & 5.0e9 & \bf{0.09} & \bf{12.9}& \bf{8.4}\\
\hline
&&&&&&
& $\Omega$=2.3h & 9.9e9 & \bf{0.18} & \bf{23.9} & \bf{16.8}\\
\hline
&&&&&&
& $\phi$=25$^o$ & 1.1e8 & 0.0002 & 0.03 & 3e-4\\
\hline
&&&&&&
& $\phi$=45$^o$ & 3.0e9 & 0.001 & 0.09 & 3e-4\\
\hline
%&&&&&&
%& 0.2/4.6h/25$^o$ & 3.0e9 & 0.057 & 8.24 &  \\
%\hline
&&&&&&
& 0.8/2.3h/45$^o$ & 1e10 & \bf{0.19} & 24.5 &\\
\hline

Dysnomia/Eris & 410 & 15.77 & 1.5 & 0.004 & 1.3e9 & 791$^\dagger$ & $e$=0.2 & 6.3e6 & 0.005 & 3.82 & \bf{2.58}\\
\hline
&&&&&&
& $e$=0.8 & 1.0e8 & \bf{0.078} & \bf{56.9} & \bf{2.58}\\
\hline
&&&&&&
& $\Omega$=6h & 1.4e9 & \bf{1.09} & \bf{413} & 0.056\\
\hline
&&&&&&
& $\Omega$=3h & 2.9e9 & \bf{2.19} & \bf{543} & 0.113\\
\hline
%&&&&&&
%& $\phi$=25$^o$ & 4.0e6 & 0.003 & 2.4 & 0.004 \\
%\hline
&&&&&&
& $\phi$=45$^o$ & 1.1e7 & 0.009 & 6.7 & 0.004\\
\hline
%&&&&&&
%& 0.2/6h/25$^o$ & 1.1e9 & \bf{0.873} & \bf{369} &  \\
%\hline
&&&&&&
& 0.8/3h/45$^o$ & 3.0e9 & \bf{2.27} & \bf{549} &\\
\hline

\hline
Pluto/Charon & 1187 & 6.387 & 1.88 & 0.052 & 2.4e10 & 363
& e=0.5 & 1.2e11 & \bf{5.0} & \bf{304} & 0.005\\
\hline
&&&&&&
& $\Omega$=2.7h & 4e12 & \bf{166} & \bf{361} & 0.026\\

\hline
Charon/Pluto & 606 & 6.387 & 1.71 & 0.012 & 2.6e9 & 873$^\dagger$
& e=0.5 & 6.3e10 & \bf{24} & \bf{838} & 0.009\\
\hline
&&&&&&
& $\Omega$=2.8h & 2e12 & \bf{760} & \bf{871} & 0.002\\
\hline
Orcus/Vanth & 458 & 9.540 & 1.5 & 0.006 & 2.0e9 & 638$\dagger$
& e=0.5 & 4.5e6 & 0.002 & 1.4 & 0.33\\
\hline

&&&&&&
& $\Omega$=3h & 2e8 & \bf{0.10} & 56 & \bf{3.1}\\
\hline

Vanth/Orcus & 189 & 9.540 & 1.5 & 0.001 & 1.1e8 & 1992$^\dagger$
& e=0.5 & 2.0e6 & 0.02 & 36 & 0.74\\
\hline

&&&&&&
& $\Omega$=3h & 8.9e7 & \bf{0.81} & \bf{890} & 0.073\\
\hline

Hi'iaka/Haumea & 160 & 49.12 & 1.5 & 4e-4 & 8.2e7 & 1911$^\dagger$
& $\Omega$=3h & 2.8e4 & $\sim$0 & 0.65 & \bf{69}\\
\hline

Namaka/Haumea & 85 & 18.27 & 1.5 & 8e-5 & 1.2e7 & 3597$^\dagger$
& $\Omega$=3h & 1.2e4 & 0.001 & 3.5 & \bf{4.6}\\
\hline

2007OR10/S255088 & 768 & $\geq$7 & 1.5 & 0.03 & 7.3e9 & 490 
& $\Omega$=3h  & 6e6  & 0.001  &  0.4 & \bf{2560} \\
\hline

1$M_{Er}$/2.5$M_{Dy}$ & 1163 & 15.77 & 2.5 & 0.089 & 5.4e10 & 154 
& $\Omega$=2.3h  & 6.3e10  & \bf{1.2}  &  \bf{83} & \bf{2.6} \\
\hline

2$M_{Er}$/2$M_{Dy}$ & 1464 & 11.0 & 2.5 & 0.14 & 1.1e11 & 122 
& $\Omega$=2.3h  & 2.7e11  & \bf{2.5}  &  \bf{87} & \bf{3.8} \\
\hline

2$M_{Er}$/2$M_{Dy}$ @ 0.8$a$ & 1464 & 7.8 & 2.5 & 0.14 & 1.1e11 & 122 
& $\Omega$=2.3h  & 1.5e12  & \bf{13.5}  &  \bf{114} & \bf{1.4} \\
\hline

\hline
\hline
\end{tabularx}
}

\normalsize

\section{Discussion}
\label{sec4}

\subsection{Subsurface Water}
\label{subwater}

While the influence of radiogenic heating alone suggests that a number of larger TNOs may possess subsurface oceans \parencite{HUSSMANN2006}, it is uncertain whether smaller bodies near the size of Charon are also capable of maintaining long lived oceans. Work done by \textcite{Desch2009694} suggests that bodies of similar sizes to Charon can only maintain subsurface oceans if the liquid layer is composed of water which contains antifreezes, or if there is an insulating undifferentiated crust that inhibits cooling. If TNO interior liquid layers do contain antifreeze, however, they may result in long lived oceans that may persist to or near the present day. Figure 11 of \textcite{Desch2009694} suggests that even bodies like Haumea and Quaoar, along with Charon, may have possessed subsurface liquid for $>$ 4 Gyr if their liquid layers had an ammonia content of $\sim$1$\%$.  
While ammonia content of such a layer could be determined by the ammonia abundance of the accreting material that formed the TNOs, an additional means of increasing antifreeze content in a liquid layer would be leaching by interactions between rock and water as heating of a body melts additional interior layers. Our calculated change in depth of a liquid layer due to tidal heating, $\Delta{}D$, may play a critical role in sequestering antifreeze to liquid layers in these bodies. Indeed, this additional depth of melt is essentially a differentiation height and would suggest that even the time duration of tidal heating is less critical if the magnitude of heating is strong enough to melt a significant layer and leach additional antifreeze. Even as the liquid layer then cools and partially solidifies, the consequently increasing concentration of antifreeze and subsequently the freezing point of the remaining liquid will be dependent on the past thermal history of the body.  Thus, in addition to cases of longer lived tidal heating that is not efficiently lost, scenarios which include short episodes of significant tidal heating can enable TNOs to remain ocean worlds to the present.

An interesting counterpoint that we have found is that suppressing the freezing temperature will affect tidal dissipation within the ice shell. This occurs in two manners: 1). The peak temperature at which viscoelastic tides are most efficient may be shifted outside of the expected range for a particular ice layer. 2). While, liquid-water tides are beyond the scope of this work, as the ocean layer increases it will reduce the viscoelastic ice depth thereby increasing its volume. This increased volume will allow for greater dissipation as our model assumes a volume fraction scale. That greater dissipation will increase heating while decreasing dampening times. This results in shorter lived tides. The balancing between anti-freezes and long-term tidally driven habitability is a complex one that deserves further study.

\subsection{Cryovolcanism \& Tectonism}
\label{cryo}

The results above also effect potential cryovolcanism. Given how close some of the bodies described in \textcite{Desch2009694} may be to still possessing interior liquid even to the present day, the contribution of even moderate tidal heating for merely hundreds of millions of years may result in a tipping point for present cryovolcanism.

An additional effect of tidal heating on TNO cryovolcanism arises from findings in previous work \parencite{Neveu201548} that suggest gas-driven (explosive) cryovolcanism is less likely on TNOs that maintain large undifferentiated crusts. Cryovolcanism is driven by the freezing of liquid or volatiles in cracks in the crust that yields greater overpressure for the remaining liquid and gas. This overpressure drives growth of the crack and potentially explosive volcanism. The density difference in states of the freezing material controls the overpressure. The optimum ratio for greater overpressure is best for pure water and poorer for certain antifreeze eutetics. Consequently, warm tidally heated and processed crusts that have sequestered antifreeze to interior liquid layers and produced water ice heavy outer layers are likely to produce the best pressure differentials for cryovolcanism. Undifferentiated crusts are also likely to inhibit cryovolcanism since they possess higher fracture toughness than pure water \parencite{Neveu201548}. Thus, tidally heated bodies that fully separate rock from water would be more amenable to cracking. Lastly, long tidal durations mean long periods where a crust is exposed to tidal stresses (even while the body was experiencing net cooling) that may form cryvolcanic vents.

In fact, using the multilayer viscoelastic methods of \textcite{HenningHurford2014}, we model the surface stress state of both Eris and Dysnomia for a range of eccentricities and shell thicknesses $d$. We assume an ice rigidity of 4$\times$10$^9$ Pa, and viscosity structure for a 10 layer ice shell with a conductive geotherm. This results in maximum tensile stresses $\sigma_{max}$ per orbit that exceed a fracture threshold of 20 kPa implied for fracturing of Europa's ice shell \parencite{Rhodenetal2010,RhodenHurford2013}. For example, $\sigma_{max}$ = 66 kPa for Eris at $e$ = 0.3 with a near-worst case ice shell of 167 km thickness, or $\sigma_{max}$ = 61 kPa for Dysnomia at the same eccentricity but $d$ = 20 km (Europa-like). We predict that these bodies will exhibit tectonism not unlike that seen on Charon. Additionally, tidally driven tectonism can be exploited to enhance cryovolcanic activity. The magnitude of dilation of the surface is in the range 1--20 m depending on $e$ and $d$, which may be distributed among any number of fractures around a global meridian to assess the magnitude of cyclic fracture openings. Moreover, tidally generated fractures can probe as deep as a few km into the body for the stresses noted above, depending on the size of the body. This preliminary portrait of surface stresses is favorable for assisting past and maybe even present cryovolcanism. 

\subsection{Future Work and Relevance to Observations}
\label{Future}

Results of this study indicate a thorough investigation of binary TNO tidal-thermal histories is warranted. Unlike Pluto-Charon, other pairs are not yet constrained to be in a dual synchronous state, opening the possibility that tidal damping, including fracturing and cryovolcanism, remains underway. In future work, we will extend our preliminary multilayer tidal calculations to report on how a time varying viscoelastic layer system with a flexible ice shell responds during secular crystallization. There is also a strong need for a detailed viscoelastic despinning analysis that includes the possibility of capture into non 1:1 spin-orbit resonances \parencite{Makarov2012}. Hybrid fluid-solid tidal models should also be considered \parencite{TylerHenningHamilton2015}. Detailed examination of the relevant eccentricity/spin phase spaces will better determine the likely origins of Pluto-Charon's current orbital state. We also highlight the strong need for hydrodyanmic simulations to help constrain post-impact parameters.

The potential that tidal heating may have lengthened post-collisional TNO subsurface oceans durations should be incorporated into targeting of these systems for observations. Besides Pluto-Charon, Eris-Dysnomia appears to be the next most likely system to have a subsurface ocean that was potentially enlarged or had its existence prolonged by tidal action. These bodies should also be subject to heightened spectroscopic scrutiny to search for relatively recent cryovolcanism markers.

The prediction that some of these systems may have bodies that have not yet damped to a synchronous rotational state suggests long-term photometric monitoring of these systems will be an important step to gleaning insight into their past history.  Given that some TNO binaries do not yet even have well-determined orbital periods (such as the 2007 OR10 system), we also emphasize the need for more basic observations to determine the current orbital states of TNO system members. Such work will be vital to bounding the thermal history and potential for subsurface liquid water on this large reservoir of possible ocean worlds.

 \bibliographystyle{model1b-num-names}

\bibliography{tidaltno}

\end{document}